\documentclass[prl,aps,amsmath,amsfonts,amssymb,reprint,floatfix]{revtex4-1}
\pdfoutput=1
\usepackage{graphicx}
\usepackage{graphics}
\usepackage{xcolor}
\usepackage{bm}
\usepackage{array}
\usepackage[percent]{overpic}
\usepackage{physics}

\begin{document}
\title{Microwave Conductivity Distinguishes Between Different d-wave States: Umklapp Scattering in Unconventional Superconductors}
\author{D. C. Cavanagh}
\email{david.cavanagh1@uqconnect.edu.au}
\affiliation{School of Mathematics and Physics, The University of Queensland, Brisbane, Queensland 4072, Australia}
\author{B. J. Powell}
\affiliation{School of Mathematics and Physics, The University of Queensland, Brisbane, Queensland 4072, Australia}

\begin{abstract}
Microwave conductivity experiments can directly measure the quasiparticle scattering rate in the superconducting state. We show that this, combined with knowledge of the Fermi surface geometry, allows one to distinguish between closely related superconducting order parameters, e.g.,  d$_{x^2-y^2}$ and d$_{xy}$ superconductivity. We benchmark this method on YBa$_2$Cu$_3$O$_{7-\delta}$ and, unsurprisingly, confirm that this is a d$_{x^2-y^2}$ superconductor. We then apply our method to $\kappa$-(BEDT-TTF)$_2$Cu[N(CN)$_2$]Br, which we discover is a d$_{xy}$ superconductor.
\end{abstract}
\maketitle

In many unconventional superconductors, the tunnelling experiments that definitively identified the superconducting gap symmetries in cuprate superconductors \cite{Wollman1993,Tsuei1994} are prohibitively difficult to perform. This presents a significant difficulty in distinguishing the form of the order parameter and therefore in understanding the microscopic origin of superconductivity, which cannot be probed directly in macroscopic experiments. Conventional probes of the superconducting gap rely on the use of extremely low temperature measurements, where the temperature dependence can be used to identify the low energy density of states \cite{Annett1990,Ketterson1999,Mineev1999}. This then gives insight into the nature of low energy excitations in the superconducting system. 

While such methods are useful in distinguishing between nodeless `s-wave' superconductivity, gaps with  line nodes, and those with point nodes, they are unable to resolve the exact form of the superconducting gap. For example, the temperature dependence of the low temperature heat capacity may identify the presence of line nodes, but in order to determine the location of such nodes on the Fermi surface, more complicated directional probes have been necessary \cite{Malone2010,Nakazawa1997}. Such experiments with directional resolution are  difficult to perform and interpret, which motivates one to discover for new probes or learn how to gain more information from existing experiments. 

The temperature dependence of the penetration depth is often measured via microwave conductivity experiments and has long been used as a probe of unconventional superconductors. The exponential suppression of low energy quasiparticles in conventional (nodeless) superconductors is evident in an activated exponential temperature dependence of the penetration depth. In contrast, the penetration depth in unconventional (nodal) superconductors exhibit a power law (often linear) temperature dependence at low temperatures \cite{Tinkham2004,Uemura1991,Hirschfeld1994,Prozorov2006}.

In this Letter, we propose a richer use of the microwave conductivity, as a more detailed probe of the superconducting gap structure. This  relies  on the ability of the microwave conductivity to accurately determine the relaxation rate of the superconducting quasiparticles, responsible for the screening of the Meissner effect. We show that, in conjunction with the knowledge of the Fermi surface geometry in the normal state, measurements of the quasiparticle relaxation rate via the penetration depth can be used to differentiate between closely related order parameters, e.g., d$_{x^2-y^2}$ and d$_{xy}$.

To make these ideas concrete we focus on two families of unconventional superconductor: the  high-temperature cuprate superconductors and organic superconductors. 
The $d_{x^2-y2}$ pairing symmetry of YBa$_2$Cu$_3$O$_{7-\delta}$ (YBCO) is long established \cite{Scalapino1995,Wollman1993,Tsuei1994}. At low temperatures the quasiparticle scattering rate varies exponentially with temperature \cite{Hosseini1999}, which results from umklapp scattering becoming an activated process in the $d_{x^2-y2}$ superconducting state \cite{Walker2000,Duffy2001}. Here we show that the experimentally observed penetration depth is incompatible with d$_{xy}$ pairing, as this would allow umklapp scattering at arbitrarily low energies. The conclusion that the pairing symmetry is $d_{x^2-y2}$ in YBCO is not new, but demonstrates the potential additional information available in microwave conductivity experiments.

Despite being one of the most studied organic superconductors, the pairing symmetry of  $\kappa$-(BEDT-TTF)$_2$Cu[N(CN)$_2$]Br ($\kappa$-Br) remains  contested \cite{Wosnitza2012, Wosnitza2003, Elsinger2000, Dion2009, Kanoda1996, Nakazawa1997, Taylor2007, Kuehlmorgen2017, Schmalian1998, Kino1998,Kondo1998, Kotegawa2002,Kuroki2002,Powell2004, Powell2005,Powell2006,Guterding2016a,Guterding2016, Watanabe2017, Zantout2018}. High resolution measurements of the quaisparticle relaxation rate in $\kappa$-Br have only been performed in recent years, and show a cubic temperature dependence, as opposed to the activated exponential seen in YBCO \cite{Milbradt2013}. 
This shows that umklapp scattering occurs at arbitrarily low energies in the superconducting state of $\kappa$-Br. We show that the only order parameter, of those discussed for $\kappa$-Br, consistent with gapless umklapp scattering has d$_{xy}$ symmetry.

\begin{figure}
	\centering
	\begin{overpic}[trim = 0mm 0mm 0mm 0mm, clip, width=0.2\textwidth]{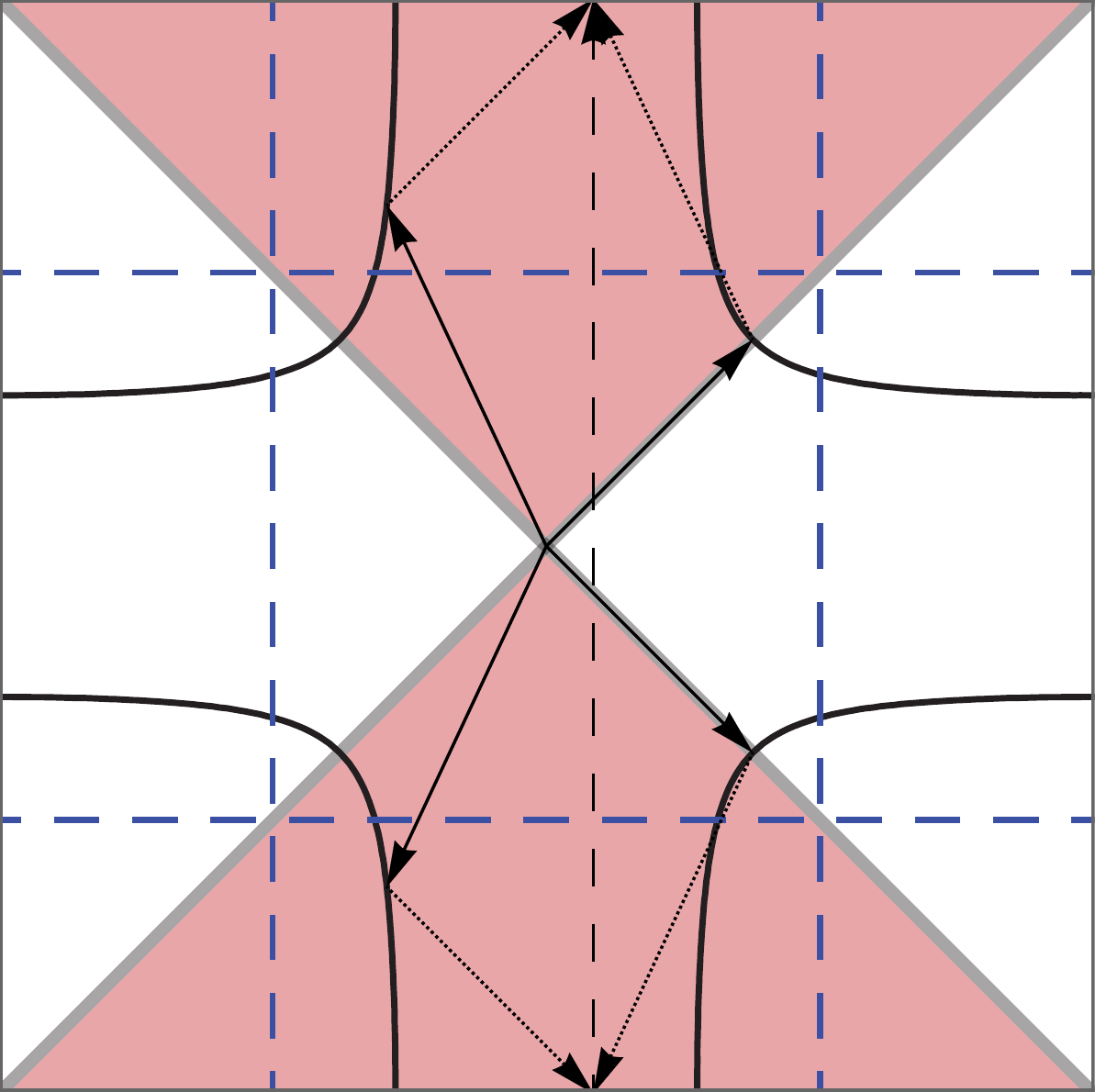} %, grid,scale=0.5,unit=1mm  
		\put (45, 63) {$\bm{k}_2$}%
		\put (64, 56) {$\bm{k}_1$}%
		\put (45, 33) {$\bm{k}_4$}%
		\put (64, 40) {$\bm{k}_3$}%
		\put (40, 80) {$\bm{G}_y$}%
		\put (-10, 100) {\bf a)}
	\end{overpic} \qquad 
	\begin{overpic}[trim = 0mm 0mm 0mm 0mm, clip, width=0.2\textwidth]{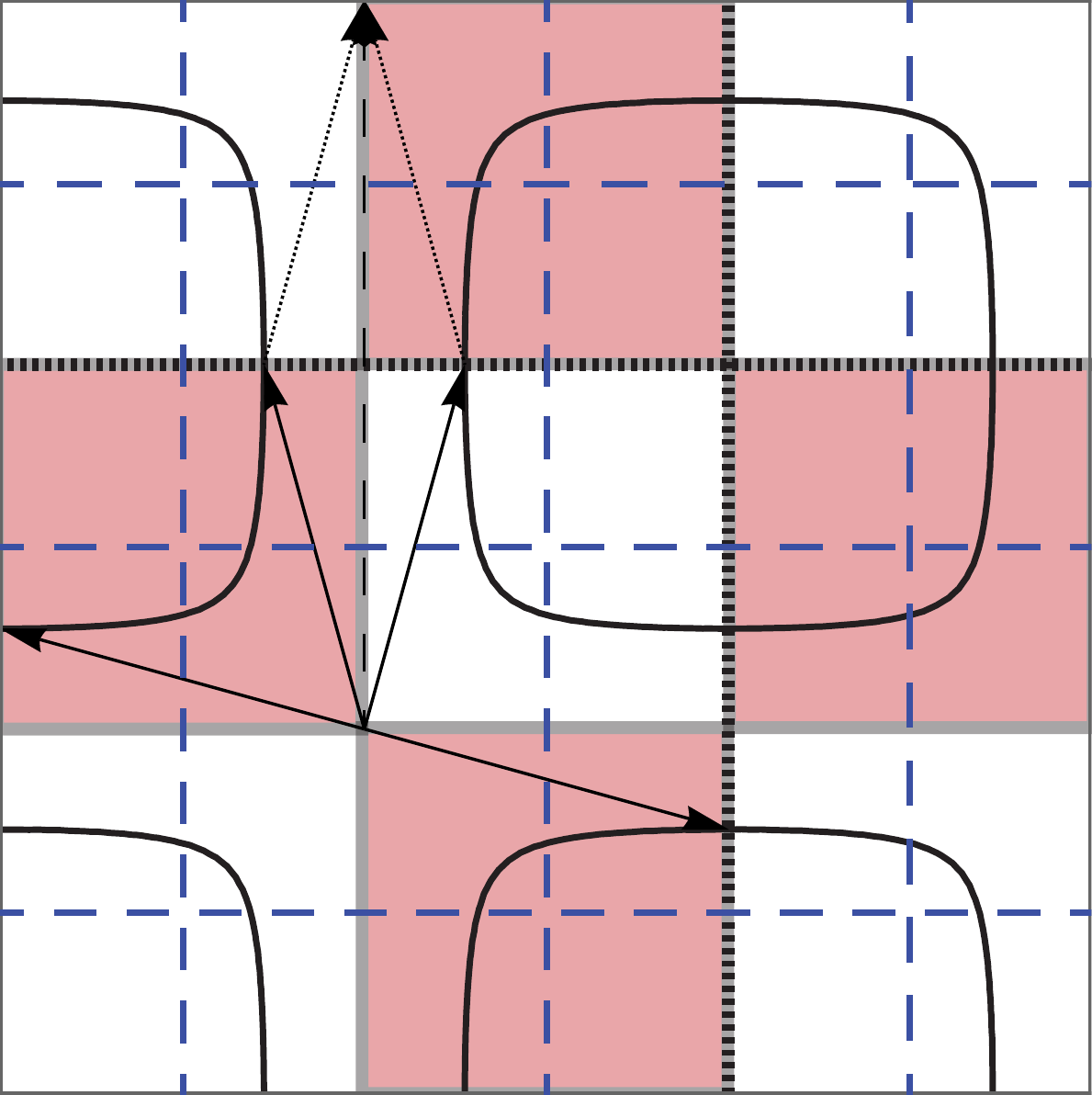} % , grid,scale=0.5,unit=1mm 
		\put (37, 42) {$\bm{k}_1$}%
		\put (28, 58) {$\bm{k}_2$}%
		\put (3, 34) {$\bm{k}_4$}%
		\put (51, 31) {$\bm{k}_3$}%
		\put (27, 70) {$\bm{G}_y$}%
		\put (-10, 100) {\bf b)}
	\end{overpic}\\
	\includegraphics[trim = 0mm 0mm 0mm 0mm, clip, width=0.4\textwidth]{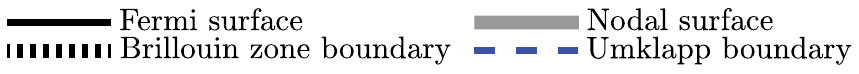}
	%	\begin{overpic}[trim = 80mm 120mm 50mm 125mm, clip, width=0.2\textwidth]{Axes_blank.pdf} %  , grid,scale=0.5,unit=1mm
	%	\put (32, 50) {$\bm{k}_y$}%
	%	\put (72, 10) {$\bm{k}_x$}%
	%	\end{overpic}
	\caption{Typical umklapp scattering process involving nodal quasiparticles in  YBCO, assuming (a) d$_{x^2-y^2}$ pairing and (b) d$_{xy}$ pairing, in this panel we plot momenta outside of the FBZ for clarity. In (a) umklapp scattering necessarily involves particles away from the nodes, leading to an exponentially activated umklapp scattering rate. In (b) umklapp scattering is possible involving only nodal quasiparticles, leading to $\tau_u^{-1}\propto T^3$. The activated temperature dependence of the relaxation rate is therefore  sufficient to distinguish between these two gap symmetries, independent of other experiments. Both panels use the extended scheme and show two BZs,  dashed arrows indicate the reciprocal lattice vectors,  dotted arrows are guides to the eye to show the total momentum before and after the scattering process. The Fermi surface is calculated from the single band model for YBCO  \cite{Duffy2001}, and the shading denotes the sign of the order parameter.}
	\label{UmkSurf_WalkerSmith}
\end{figure}
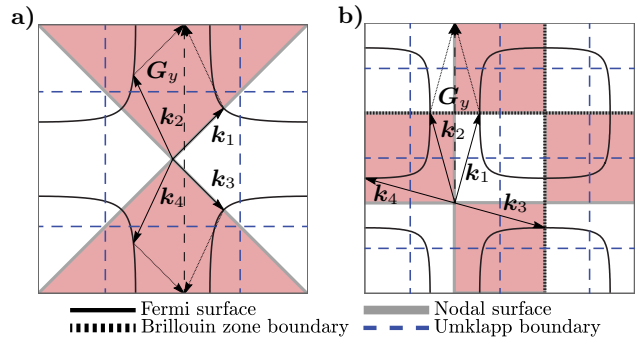

%Trim
The current is proportional to the total momentum carried by the electronic quasiparticless and therefore cannot be relaxed by elastic processes. Current relaxation due to elastic electron-electron interactions requires the presence of some mechanism for the loss of momentum. The most significant mechanism for such relaxation is umklapp scattering. Where the initial ($\bm{k}_1$, $\bm{k}_2$) and final ($\bm{k}_3$, $\bm{k}_4$) momentum states satisfy
\begin{eqnarray}
\bm{k}_1+\bm{k}_2-\bm{k}_3-\bm{k}_4&=&\pm\bm{G}_j.\label{x_UmkCond}
\end{eqnarray}
and $\bm{G}_j$ is a reciprocal lattice vector. Umklapp scattering transfers momentum to the lattice allowing the current to relax due entirely to elastic electron-electron scattering \cite{Yamada1989, Maebashi1997, Maebashi1998, Walker2000, Duffy2001, Rosch2005,Rosch2006,Pal2012}. 
For a simple Fermi liquid, with a circular Fermi surface, umklapp scattering requires  $k_F\geq\left|\bm{G}_j/4\right|$. For non-trivial band structures, it is natural to define an `umklapp boundary' of $\left|\bm{k}_j\right|\geq\bm{G}_j/4$ in the first Brillouin zone (FBZ), see Fig. \ref{UmkSurf_WalkerSmith}. Umklapp scattering can occur between states on different sides the umklapp boundary \cite{Walker2000,Rosch2005}. 

In a normal metal, as long as the umklapp condition Eq. (\ref{x_UmkCond}), is satisfied for some points on the Fermi surface, the current can relax entirely due to electron-electron scattering, yielding the well-known quadratic temperature dependence of the resistance \cite{Yamada1989}. The scattering rate due to umklapp processes, which relax the current, and the total electron-electron scattering rate, including elastic processes, differ only by an overall factor, due to the reduced phase space available for umklapp processes \cite{Yamada1989,Rosch2005}. For a superconductor with a nodal gap function, however, the structure of the superconducting gap reduces the phase space available for umklapp scattering. The total scattering rate and current relaxation rate can therefore differ more dramatically than in a normal metal.

Walker and Smith \cite{Walker2000} first addressed this possibility theoretically, after experiments on YBCO found a current relaxation rate with an activated exponential temperature dependence, rather than the cubic dependence found for the total scattering rate \cite{Hosseini1999}. The central argument of their theory is that, because the nodes of the gap on the Fermi surface in YBCO do not satisfy the umklapp condition, Eq. (\ref{x_UmkCond}), any umklapp scattering process must necessarily involve quasiparticles away from the nodes. They showed geometrically that in YBCO with d$_{x^2-y^2}$ superconductivity, no umklapp process is possible involving only nodal quasiparticles (see Fig. \ref{UmkSurf_WalkerSmith}a). This necessarily leads to a relaxation rate for umklapp processes involving two nodal quasiparticles of $\tau^{-1}_{u}\propto T^2f\left(\Delta_U\right)\bar{f}\left(\Delta_U\right)\propto T^2\exp\left(-\Delta_U/k_BT\right)$ where the umklapp $\Delta_U$ is the minimum energy of the states to which non-nodal quasiparticles can umklapp scatter. More generally, we find, that $\Delta_U$ is the minimum energy of any four states that satisfy Eq. (\ref{x_UmkCond}). 
%This umklapp gap is responsible for the exponential temperature dependence of the quasiparticle scattering rate in YBCO \cite{Hosseini1999}. 
The total scattering rate, in contrast, can be found from power counting of scattering processes involving the nodal quasiparticles to be $\tau^{-1}\propto T^3$. Both scattering rates were later reproduced by  numerical calculations based on the random phase approximation (RPA) \cite{Duffy2001}.

The existence of these two distinct temperature dependences for the total and umklapp scattering rates is in general expected in nodal superconductors, with the energy scale $\Delta_U$ set by the geometry of the gap function and the nodal placement.
In this Letter we identify a crucial exception to this rule: if the nodes exactly satisfy the umklapp condition, Eq. (\ref{x_UmkCond}),  then the umklapp scattering rate will be dominated by the contribution due to the low energy nodal quasiparticles, and will vary cubically, rather than exponentially, with temperature. Crucially, we show that a nodal placement satisfying the umklapp condition is \textit{not} an exotic occurrence requiring fine tuning, rather, this is required for certain combinations of pairing symmetry and Fermi surface geometry. A key example below will be d$_{xy}$ pairing on a Fermi surface that crosses the boundary of the FBZ, which allows umklapp scattering between quasiparticles exactly at the nodes.

As a demonstrative example, we first consider an alternative d$_{xy}$ state in YBCO. In Fig. \ref{UmkSurf_WalkerSmith}, we sketch the possible momentum configurations for umklapp scattering involving nodal quasiparticles in a realistic model of YBCO for the two d-wave gaps. For a d$_{x^2-y^2}$ gap, an umklapp scattering process involving quaiparticles at the nodes, inside the umklapp boundary, must also involve states outside the boundary and away from the nodes, Fig. \ref{UmkSurf_WalkerSmith}a, leading to an exponential temperature dependence. In the d$_{xy}$ case, however,  Fig. \ref{UmkSurf_WalkerSmith}b, there exists an electron configuration for which the quasiparticles at the nodes (on the FBZ boundary) contribute to the umklapp scattering, with no umklapp gap, giving a cubic temperature dependence. This example suggests the possibility of using such measurements as a more direct probe of the detailed form of the superconducting gap than has been previously considered. Any insight gained from these measurements, however, requires a detailed understanding of the underlying normal state Fermi surface.

$\kappa$-Br  provides an important opportunity to use the quasiparticle scattering rate as a probe of superconducting gap symmetry. The normal state properties have been studied in great detail \cite{Caulfield1994,Kino1995,Kino1996,McKenzie1998,Merino2000b,Koretsune2014,Guterding2015,Guterding2016a}, and  the quasiparticle relaxation rate has been measured: it varies cubically with temperature \cite{Milbradt2013}. 

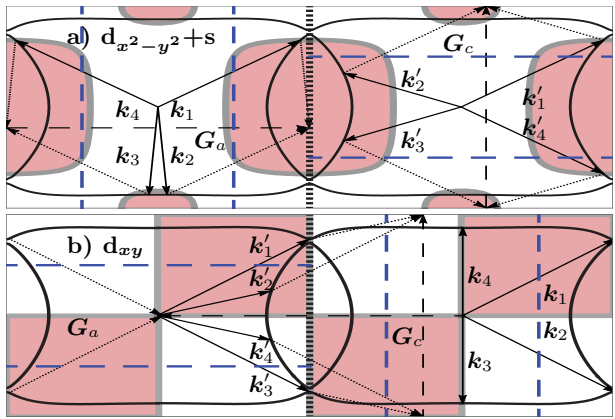
\begin{figure}
	\centering
	\begin{tabular}{c}
		\begin{overpic}[trim = 0mm 0mm 0mm 0mm, clip, width=0.45\textwidth , height=0.15\textwidth ]{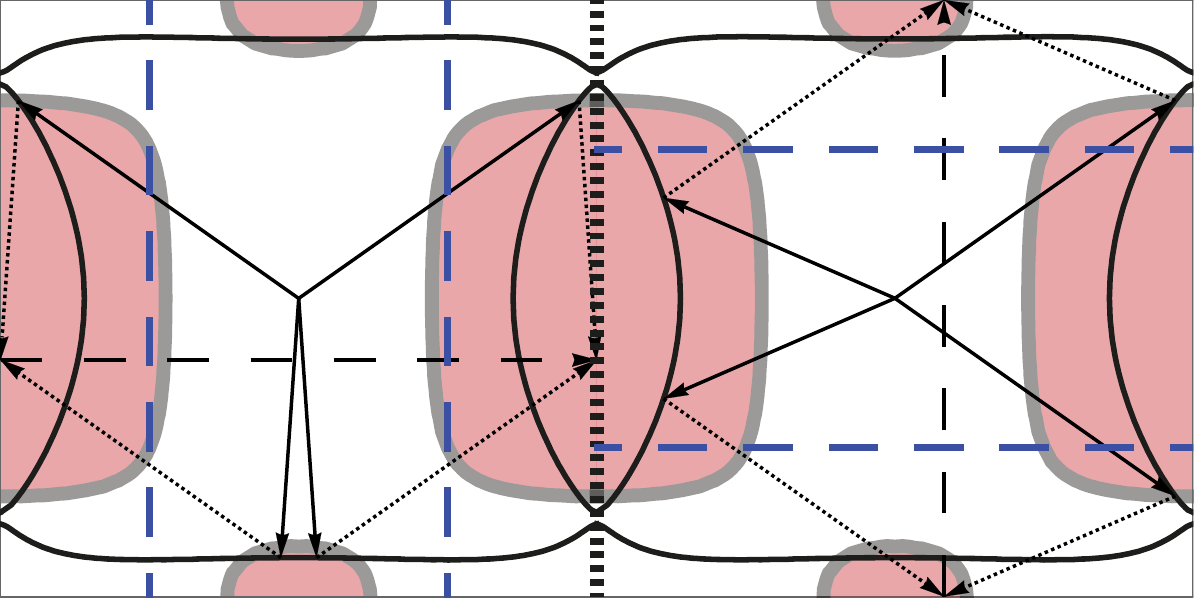} %, grid,scale=0.5,unit=1mm 
			\put (27, 7.5){$\bm{k}_2$}%
			\put (27, 15) {$\bm{k}_1$}%
			\put (18, 15)  {$\bm{k}_4$}%
			\put (18, 7.5) {$\bm{k}_3$}%
			\put (31, 10) {$\bm{G}_a$}%
			\put (65, 20) {$\bm{k}'_2$}%
			\put (85, 17) {$\bm{k}'_1$}%
			\put (85, 12) {$\bm{k}'_4$}%
			\put (65, 10) {$\bm{k}'_3$}%
			\put (72, 26) {$\bm{G}_c$}%
			\put (10, 27)  {\bf a) d${\bm{_{x^2-y^2}+}}$s}%15, 40
		\end{overpic}
		\\	
		\begin{overpic}[trim = 0mm 0mm 0mm 0mm, clip, width=0.45\textwidth , height=0.15\textwidth]{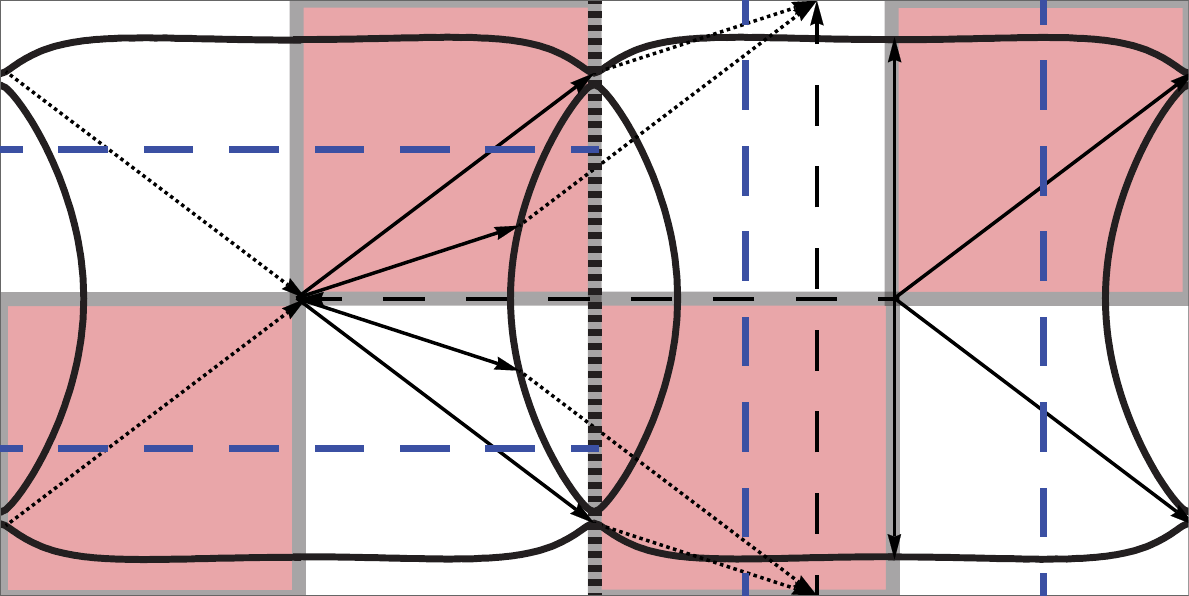} % , grid,scale=0.5,unit=1mm 
			\put (88.5, 13){$\bm{k}_2$}%
			\put (88.5, 19) {$\bm{k}_1$}%
			\put (76, 22)  {$\bm{k}_4$}%
			\put (76, 8) {$\bm{k}_3$}%
			\put (10, 13.5) {$\bm{G}_a$}%
			\put (40, 21.5) {$\bm{k}'_2$}%
			\put (40, 27) {$\bm{k}'_1$}%
			\put (40, 9.5) {$\bm{k}'_4$}%
			\put (40, 4) {$\bm{k}'_3$}%
			\put (63.5, 12) {$\bm{G}_c$}%
			\put (10, 27) {\bf b) d$\bm{_{xy}}$}%20, 7.5
		\end{overpic} 
%	\\ 
%		\includegraphics[trim = 0mm 278mm 160mm 0mm, clip, width=0.25\textwidth]{KEY_BZ.pdf}  %\begin{overpic}[trim = 80mm 120mm 50mm 125mm, clip, width=0.2\textwidth]{Axes_blank.pdf} %  , grid,scale=0.5,unit=1mm
		%	\put (32, 50) {$\bm{k}_c$}%
		%	\put (72, 10) {$\bm{k}_a$}%
		%	\end{overpic}
		% & \multicolumn{2}{*}{\includegraphics[trim = 0mm 278mm 160mm 0mm, clip, width=0.25\textwidth]{./Figs/KEY_BZ.pdf}} & %\multicolumn{2}{*}{\begin{overpic}[trim = 80mm 120mm 50mm 125mm, clip, width=0.2\textwidth]{./Figs/Axes_blank.pdf} %  , grid,scale=0.5,unit=1mm
		%	\put (32, 50) {$\bm{k}_c$}%
		%	\put (72, 10) {$\bm{k}_a$}%
		%	\end{overpic}}
	\end{tabular}
	\caption{Momentum configurations for umklapp processes involving nodal quasiparticles in $\kappa$-Br for (a) d$_{x^2-y^2}+$s pairing and (b) d$_{xy}$ pairing. For d$_{x^2-y^2}+$s pairing umklapp scattering must involve particles away from the nodes, leading to an exponentially activated umklapp scattering rate. For d$_{xy}$ pairing umklapp scattering involving only nodal quasiparticles is possible in the $c$ direction, but not the $a$ direction. Therefore at low temperatures umklapp scattering in the $c$ direction dominates leading to $\tau_u^{-1}\propto T^3$. The activated temperature dependence of the relaxation rate is therefore  sufficient to distinguish between these two gap symmetries, independent of other experiments. The gap in (b) uses the parameterization of Guterding \emph{et al}. \cite{Guterding2016,Guterding2016a}. The Fermi surface is calculated from the `monomer' model of $\kappa$-Br  \cite{Koretsune2014}, and the shading denotes the sign of the order parameter. Only the umklapp boundary relevant to scattering in a given direction is shown.}
	\label{K_umk}
\end{figure}

Both YBCO and $\kappa$-Br have $D_{2h}$ point group symmetries. In YBCO this is due to a small orthorhombic distortion, which does not change the analysis above in any significant way. In $\kappa$-Br the lattice is far from tetragonal and this has important consequences for the analysis of the superconductivity \cite{Powell2006} and the quasiparticle scattering rate. 
(i) Both YBCO and $\kappa$-Br form layered structure: the layers lie in the $a$-$b$ in YBCO and the $a$-$c$ plane in $\kappa$-Br. We  will adopt a labeling convention where the $x$ and $y$ axes are considered parallel to the $a$ and $c$ directions, respectively. However, one should note that, particularly in the theoretical literature,  superconducting order parameters are often  defined in coordinated systems rotated 45$^\circ$ from the crystal axes in $\kappa$-Br. In this basis, the `d$_{xy}$' and `d$_{x^2-y^2}$' labels are reversed. 
(ii) Unlike YBCO, the Fermi surface of $\kappa$-Br is strongly anisotropic, and as such the umklapp scattering along each of the two ($a$ and $c$) crystal axes must be considered independently, as shown in Fig. \ref{K_umk}. 

%%^^Maybe shorten?

Three different gap symmetries have been proposed in  $\kappa$-Br: 
a full gapped s-wave order parameter \cite{Elsinger2000}; 
d$_{xy}$ pairing, Fig. \ref{K_umk}a, where both the presence and the placement of nodes are required by symmetry \cite{Kondo1998,Kino1998,Powell2005,Powell2006,Powell2006a}; 
and a $d_{x^2-y^2}+s$ order parameter \cite{Powell2007,Watanabe2017,Guterding2016,Guterding2016a,Kuehlmorgen2017}, Fig. \ref{K_umk}b. In the $D_{2h}$ point group both the d$_{x^2-y^2}$ and $s$ pairing channels transform according to the trival ($A_{1g}$) irreducible representation, therefore any nodes are formally accidental, but can arise provided the d$_{x^2-y^2}$ component is larger than the s-wave component. 
 
An s-wave order parameter necessarily leads to an activated quasiparticle scattering rate, which is inconsistent with the microwave conductivity measurements in $\kappa$-Br. We display possible umklapp scattering processes, in both the $a$ and $c$ directions, involving nodal quasiparticles for both d$_{xy}$ and d$_{x^2-y^2}+$s in Fig. \ref{K_umk}. It is  clear that umklapp scattering processes  involving only nodal quasiparticles exist for a d$_{xy}$ pairing, Fig. \ref{K_umk}a, for transport in the $c$ direction, but not in the $a$ direction [i.e., only for ${\bm G}_j={\bm G}_c$ in Eq. (\ref{x_UmkCond})]. Nevertheless, at low temperatures contribution to quasiparticle scattering in the $c$ direction shorts out the contribution in the $a$ direction, leading to a cubic temperature dependence is expected for d$_{xy}$ pairing, consistent with experiment \cite{Milbradt2013}. In contrast for a d$_{x^2-y^2}+$s gap there is no set of nodes that satisfies the umklapp condition, Eq. (\ref{x_UmkCond}), Fig. \ref{K_umk}. Thus, for d$_{x^2-y^2}+$s pairing one expects the quasiparticle scattering rate to be exponentially activated -- in clear contradiction to experiment \cite{Milbradt2013}.

%The Fermi golden rule scattering rate in a superconductor is given by
%%\begin{widetext}
%\begin{eqnarray}
%\tau^{-1}\left(\bm{k}_1\right)&=&\sum\limits_{\bm{k}_2,\bm{k}_3,\bm{k}_4,j}| \tilde{V}_{\left\lbrace\bm{k}_i\right\rbrace}|^2 f\left(E_{\bm{k}_2}\right)\bar{f}\left(E_{\bm{k}_3}\right)\bar{f}\left(E_{\bm{k}_4}\right) \notag \\
%&&\times\delta\left(E_{\bm{k}_1}+E_{\bm{k}_2}-E_{\bm{k}_3}-E_{\bm{k}_4}\right) \notag \\
%&&\times\delta\left(\bm{k}_1+\bm{k}_2-\bm{k}_3-\bm{k}_4+n_j\bm{G}_j\right),\label{SCrate}
%\end{eqnarray}
%%\end{widetext}
%where $E_{\bm{k}}=\sqrt{\xi^2_{\bm{k}}+\Delta^2_{\bm{k}}}$ is the quasiparticle energy, defined in terms of the electron dispersion $\xi_{\bm{k}}$ and the superconducting gap $\Delta_{\bm{k}}$, $f\left(E\right)$ is the Fermi-Dirac distribution function [$\bar{f}\left(E\right)=1-f\left(E\right)$], the scattering potential $\tilde{V}_{\left\lbrace\bm{k}_i\right\rbrace}$ takes quasiparticle coherence factors into account, and $n_j=1$ ($n_j=0$) for umklapp (normal) scattering in the $j$ direction. 

This analysis strongly suggests that the superconductivity in $\kappa$-Br occurs in the d$_{xy}$ channel. But a little care is required. The temperature dependences discussed so far apply only as $T\rightarrow0$. Therefore, before reaching a firm conclusion, one needs to understand how large $\Delta_U$ is and hence how the quasiparticle scattering behaves at higher temperatures. To investigate this we  numerically calculated the Fermi golden rule scattering rate,
%\begin{widetext}
\begin{eqnarray}
\tau^{-1}\left(\bm{k}_1\right)&=&\sum\limits_{\bm{k}_2,\bm{k}_3,\bm{k}_4,j}| \tilde{V}_{\left\lbrace\bm{k}_i\right\rbrace}|^2 f\left(E_{\bm{k}_2}\right)\bar{f}\left(E_{\bm{k}_3}\right)\bar{f}\left(E_{\bm{k}_4}\right) \notag \\
&&\times\delta\left(E_{\bm{k}_1}+E_{\bm{k}_2}-E_{\bm{k}_3}-E_{\bm{k}_4}\right) \notag \\
&&\times\delta\left(\bm{k}_1+\bm{k}_2-\bm{k}_3-\bm{k}_4+n_j\bm{G}_j\right),\label{SCrate}
\end{eqnarray}
%\end{widetext}
where $E_{\bm{k}}=\sqrt{\xi^2_{\bm{k}}+\Delta^2_{\bm{k}}}$ is the quasiparticle energy, defined in terms of the electron dispersion $\xi_{\bm{k}}$ and the superconducting gap $\Delta_{\bm{k}}$, $f\left(E\right)$ is the Fermi-Dirac distribution function [$\bar{f}\left(E\right)=1-f\left(E\right)$], the scattering potential $\tilde{V}_{\left\lbrace\bm{k}_i\right\rbrace}$ takes quasiparticle coherence factors into account, and $n_j=1$ ($n_j=0$) for umklapp (normal) scattering in the $j$ direction.  We take the scattering potential potential to be of the RPA form, which has been shown to accurately reproduce the measures quasiparticle scattering rate in YBCO  \cite{Duffy2001}. The temperature dependence of the superconducting gap is assumed to follow a strong-coupling BCS form: $\Delta(T)=(5k_BT_c/2)\tanh[3\sqrt{(T_c/T)-1}]$, with parameters the overall magnitude and temperature dependence in agreement with experiment \cite{Milbradt2013}. 

For concreteness we considerer only the simplest d$_{xy}$ order parameter: $\Delta_{\bm{k}} \propto \sin\left(k_a\right)\sin\left(k_c\right)$ where $k_a$ and $k_c$ are the crystal momentum components along the $a$ and $c$ crystal axes. 
d$_{x^2-y^2}+$s order parameters have an additional freedom, the degree of mixing of the d$_{x^2-y^2}$ and $s$ components. To avoid treating this as a free parameter we take the recent parameterization from fits to scanning tunnelling spectra, which also includes an extended-s component: $\Delta_{\bm{k}}\propto c_{s_1}\left[\cos\left(k_a\right)+\cos\left(k_c\right)\right]+c_{s_2}\cos\left(k_a\right)\cos\left(k_c\right)+c_{d}\left[\cos\left(k_a\right)-\cos\left(k_c\right)\right]$ \cite{Guterding2016}. 
%order parameter Likewise our numerics focus on 
% Guterding \emph{et al}, with eight accidental nodes and a complicated form, $\Delta_{\bm{k}}\propto c_{s_1}\cos\left(k_x\right)\cos\left(k_y\right)+c_{s_2}\left[\cos\left(2k_x\right)+\cos\left(2k_y\right)\right]+c_{d}\sin\left(k_x\right)\sin\left(k_y\right)$, which is parametrised by a fit to scanning tunnelling spectra  \cite{Guterding2016,Guterding2016,Zantout2018}. 

In order to efficiently evaluate the integrals required to calculate the relaxation rate involving low energy quasiparticles, the quasiparticle energy $E_{k}$ was first calculated on a two dimensional grid with $2 \times 10^5$ sites per dimension. For each temperature a Monte Carlo approach was used to select a subset of these points. We calculated $\partial f(\omega)/\partial\omega$ which is peaked at low energies with a temperature-dependent width, and retaining only those points for which an appropriately normalized random number was less than $\partial f(\omega)/\partial\omega$. The resulting adaptive mesh was then used to perform the integrals required to calculate the scattering rate as described in \cite{Duffy2001}. All calculations have been performed using $\omega = 0.005t$ and broadening the $\delta$-functions to Lorentzians of width $0.0005t$. Our results vary only weakly with the bare interaction strength, $U$, and we report here only the results in the weakly interacting limit.

\begin{figure}
		\centering
	\begin{overpic}[trim = 0mm 0mm 0mm 0mm, clip, width=0.475\textwidth]{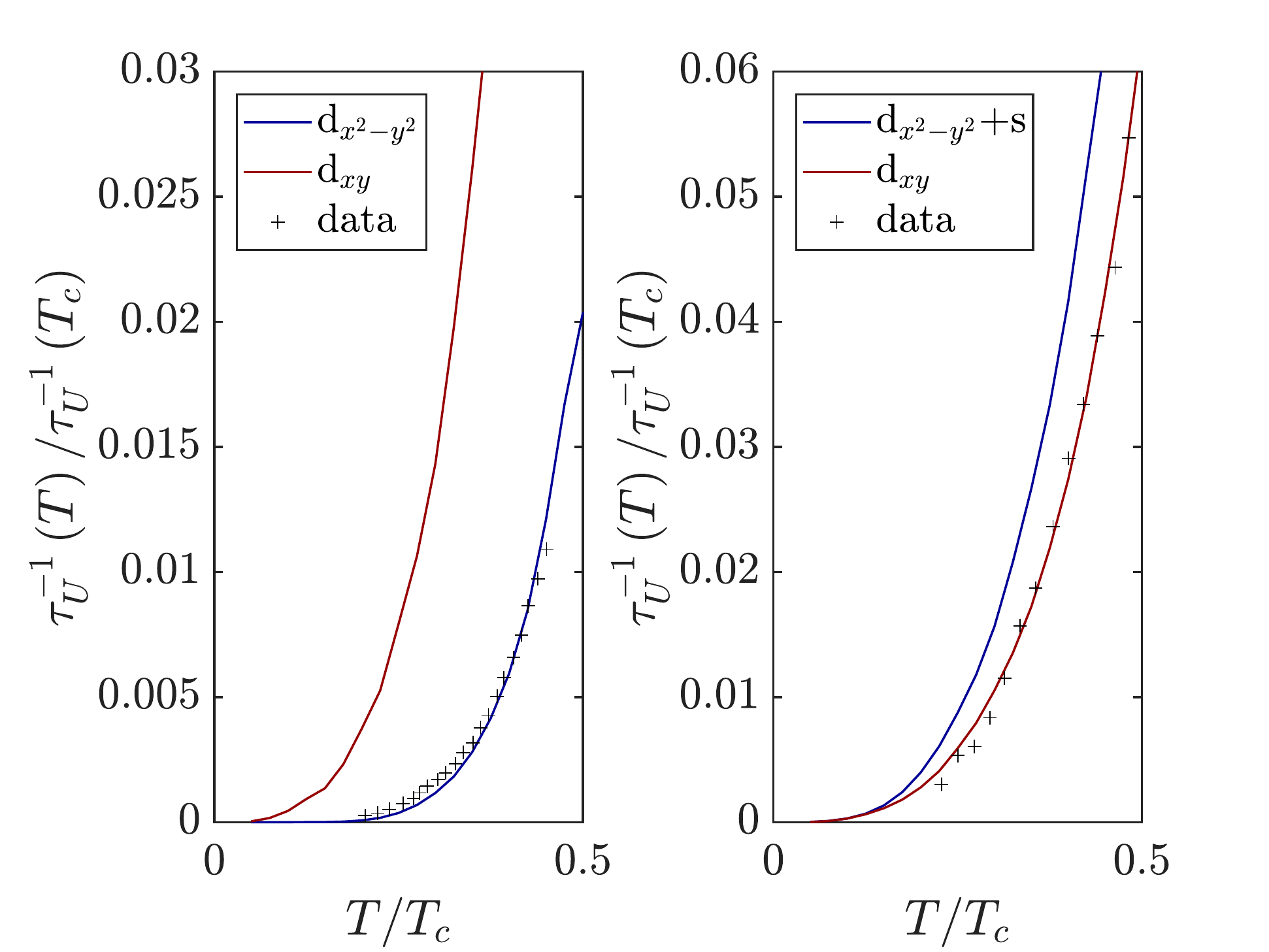} % , grid,scale=0.5,unit=1mm  
	\put (18, 27) {\includegraphics[trim = 0mm 0mm 0mm 0mm, clip, width=0.07\textwidth]{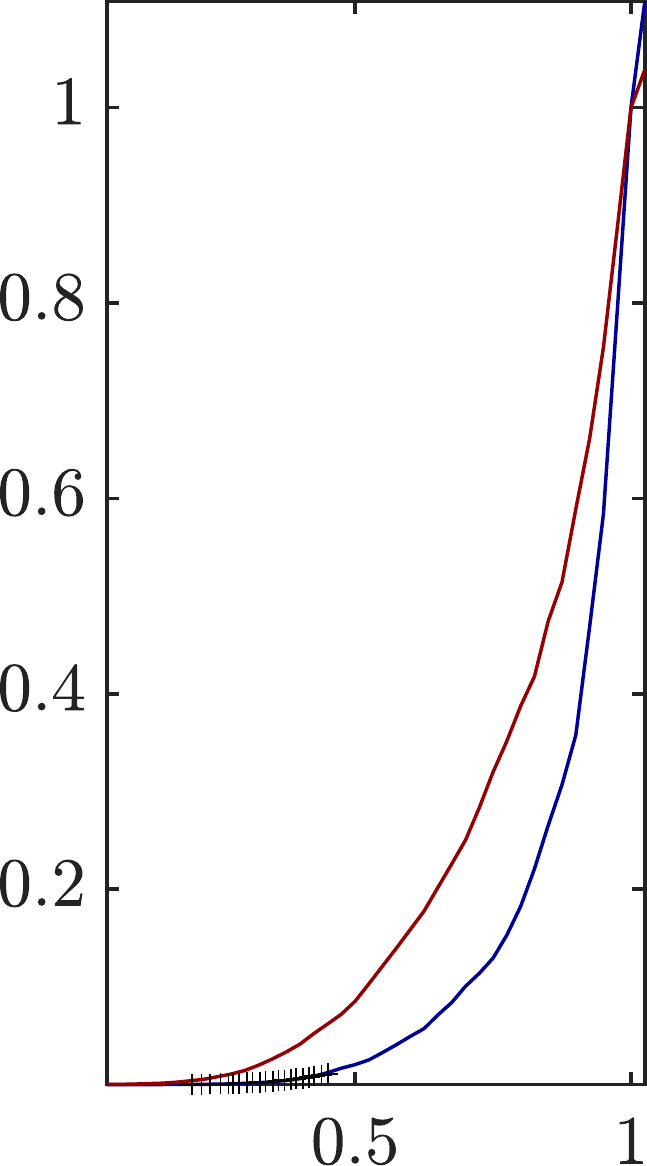}}%
	\put (63, 27) {\includegraphics[trim = 0mm 0mm 0mm 0mm, clip, width=0.07\textwidth]{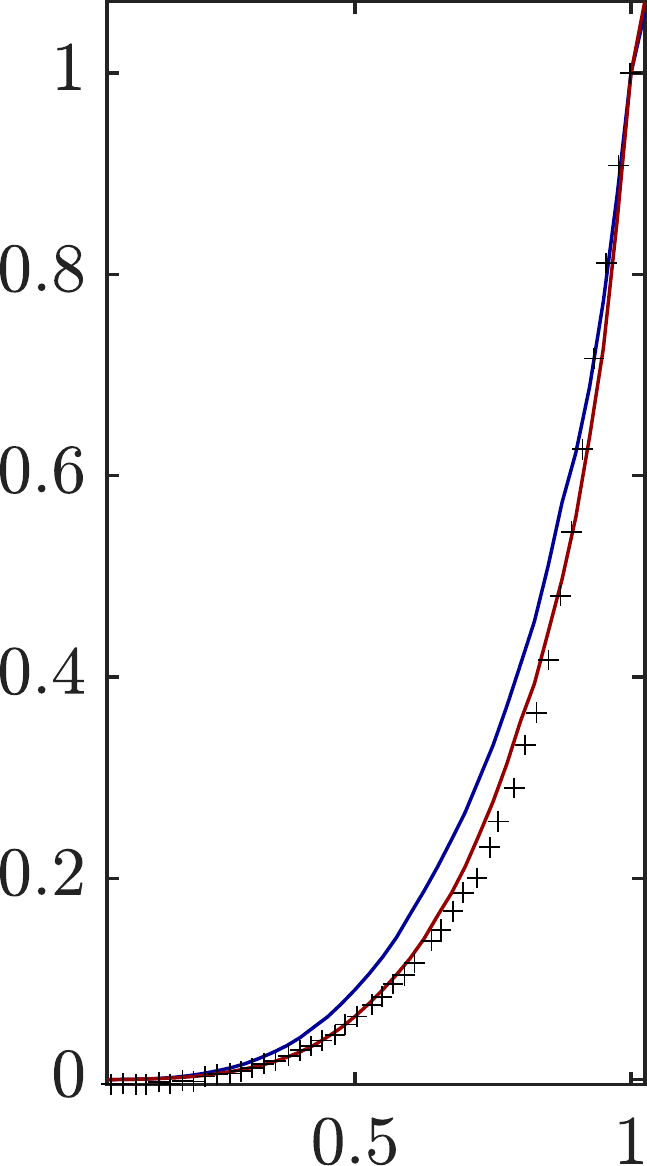}}%
	\end{overpic}
	\caption{Comparison of the calculated quasiparticle relaxation rate for the different order parameters to experimental data for (a) YBCO \cite{Hosseini1999} and (b) $\kappa$-Br \cite{Milbradt2013}. In both cases the experimental data is described accurately by only one of the order parameters: d$_{x^2-y^2}$ for YBCO and d$_{xy}$ for $\kappa$-Br.  }
		\label{Numerical}
\end{figure}

We compare the calculated umklapp scattering rates for each pairing symmetry with the Fermi surfaces for YBCO and $\kappa$-Br to the relevant experimental data \cite{Hosseini1999,Milbradt2013} in Fig. \ref{Numerical}. 
In YBCO the temperature dependence of the relaxation rate strongly differentiates between the two gap functions, with a clear cubic temperature dependence seen for the d$_{xy}$ gap, while the d$_{x^2-y^2}$ gap shows the exponential temperature dependence observed experimentally. Thus, the uncontroversial conclusion that this is a d$_{x^2-y^2}$ superconductor follows safely from this experiment alone.
For $\kappa$-Br the $\Delta_U$ for d$_{x^2-y^2}+$s pairing is somewhat smaller than the umklapp gap in YBCO. Therefore, the results are less distinct at higher temperatures, but as $T\rightarrow 0$ it is clear that the cubic temperature dependence arising from the d$_{xy}$ gap gives is in much better agreement with experiment than the exponential temperature dependence from d$_{x^2-y^2}+$s pairing. This is compelling evidence  that $\kappa$-Br is a d$_{xy}$ superconductor.

An important distinction should be made between the d$_{x^2-y^2}$ gap in YBCO and the d$_{x^2-y^2}+$s gap in $\kappa$-Br. In both cases, the umklapp processes are exponentially suppressed, but in the latter case the predicted scattering rate exceeds the $T^3$ rate for the d$_{xy}$ gap at experimentally relevant temperatures. This is due in part to the significantly smaller $\Delta_U$ for the d$_{x^2-y^2}+$s gap in $\kappa$-Br. But, more importantly, the nodes are accidental in $\kappa$-Br, rather than symmetry required as in YBCO gaps. In this case, the variation of the gap in the direction perpendicular to the Fermi surface is nonzero, resulting in a scattering rate that varies as $T^3\exp\left(-\Delta_U/k_BT\right)$, rather than $T^2\exp\left(-\Delta_U/k_BT\right)$. Finally, a large non-universal prefactor also enhances scattering in the d$_{x^2-y^2}+$s gap in $\kappa$-Br.

We must also note a technical loophole in the above argument.  For a superconducting gap with accidental nodes, there is no restriction on the nodal placement on the Fermi surface. It is possible to fine tune  the gap  d$_{x^2-y^2}+$s case to give a nodal placement satisfying Eq. (\ref{x_UmkCond}). We find that the best fit to the experimental scattering rate data is given if the s-wave component is negligibly small. For a pure d$_{x^2-y^2}$ gap ($\Delta_{\bf k} \propto \cos k_a - \cos k_c$) the calculated umklapp scattering rate is numerically indistinguishable from the d$_{xy}$ case. However, such a gap is theoretically extremely unlikely as, if allowed by symmetry, the system should always be able to lower its energy by including a s-wave admixture to the gap. This conclusion is supported by numerous microscopic calculations \cite{Powell2007,Watanabe2017,Guterding2016,Guterding2016a}, which find a significant s-wave component. Similarly, previous interpretations of other experiments in terms of a d$_{x^2-y^2}+$s gap require a sizable s-wave component.
Finally, the lack of a Hebel-Slichter peak in the nuclear magnetic resonance \cite{Kanoda1996} suggests that accidental nodes are unlikely \cite{Cavanagh2018}. %Additionally, in the case of the Guterding gap, such fine tuning to satisfy the requirements of the fitting of the parameters to experiment \cite{Guterding2016}.

%\footnote{For allowed scattering processes, the three momentum integrations each give a factor of $E^2$, while the energy and momentum conserving delta functions contribution $E^{-1}$ and $E^{-2}$, respectively, leading to the overall result $\tau\sim E^3 \sim \left(k_BT\right)^3$. The scaling argument here was confirmed numerically for the total scattering rate of YBCO, with symmetry required nodes, by Duffy \emph{et al.} \cite{Duffy2001}.}

In $\kappa$-Br, the anisotropy of the relaxation rate could provide a further test of the superconducting gap. For d$_{xy}$ pairing, the overall relaxation rate is  dominated by the contribution in the $a$ direction, as it is only in this direction that there is umklapp scattering involving only nodal quasiparticles. Thus, we predict that, a directional measurement of the relaxation rate will show an exponential temperature dependence along the $c$ direction and a cubic dependence in the $a$ direction. %Such measurements are technically demanding, however, and have not been performed as yet on this material. %Thus we propose such a measurement as an additional confirmation of the superconducting gap structure.

\begin{acknowledgements}
	We thank David Broun for helpful conversations. This work was supported by the Australian Research Council (Grant No. DP180101483) and by and an Australian Government Research Training Program Scholarship.
\end{acknowledgements}

\bibliographystyle{apsrev4-1}
\bibliography{Ongoing_Bib}

\end{document}